\documentclass[aps,preprint,showpacs]{revtex4}
\begin{document}

\preprint{{DOE/ER/40736-339}\cr{UMPP\#05-043}}

\count255=\time\divide\count255 by 60 \xdef\hourmin{\number\count255}
  \multiply\count255 by-60\advance\count255 by\time
 \xdef\hourmin{\hourmin:\ifnum\count255<10 0\fi\the\count255}

\newcommand{\xbf}[1]{\mbox{\boldmath $ #1 $}}

\newcommand{\sixj}[6]{\mbox{$\left\{ \begin{array}{ccc} {#1} & {#2} &
{#3} \\ {#4} & {#5} & {#6} \end{array} \right\}$}}

\newcommand{\threej}[6]{\mbox{$\left( \begin{array}{ccc} {#1} & {#2} &
{#3} \\ {#4} & {#5} & {#6} \end{array} \right)$}}

\title{SU(3) Baryon Resonance Multiplets in Large $N_c$ QCD}

\author{Thomas D. Cohen}
\email{cohen@physics.umd.edu}

\affiliation{Department of Physics, University of Maryland, College
Park, MD 20742-4111}

\author{Richard F. Lebed}
\email{Richard.Lebed@asu.edu}

\affiliation{Department of Physics and Astronomy, Arizona State
University, Tempe, AZ 85287-1504}

\date{April, 2005}

\begin{abstract}
We extend the recently developed treatment of baryon resonances in
large $N_c$ QCD to describe resonance multiplets collected according
to the SU(3) flavor symmetry that includes strange quarks.  As an
illustration we enumerate the SU(3) partners of a hypothetical $J^P \!
= \! {\frac 1 2}^{\pm}$ resonance in the SU(3) representation that
reduces to ${\overline {\bf 10}}$ when $N_c \! = \! 3$, and reproduce
results hitherto obtained only in the context of a large $N_c$ quark
picture. While these specific quantum numbers represent one favored
set for the possible pentaquark state $\Theta^+ (1540)$, the method is
applicable to baryon resonances with any quantum numbers.
\end{abstract}


\maketitle

\section{Introduction} \label{intro}

A recent series of papers~\cite{CL,CLpenta,CLSU3} by a collaboration
led by the current authors shows how baryon resonances may be properly
treated in the context of large $N_c$ QCD.  For reviews of this
literature, see Ref.~\cite{review}.

In the standard treatment of baryons at arbitrary $N_c$, the
ground-state spin-flavor multiplet is taken to be the completely
symmetric $N_c$-tableau representation, which is the analog to the
SU(6) {\bf 56}.  Notationally, we denote such arbitrary-$N_c$
generalizations of $N_c \! = \! 3$ representations with quotes, as in
``{\bf 56}''.  It should be pointed out that the spin-flavor symmetry
of the ground state is not a rigorous result following directly from
manipulations of the QCD action, but rather an assumption whose
phenomenological predictions concur with all available experimental
measurements.  Turning this result around, one can show that the
successes of the old baryon SU(6) spin-flavor symmetry, such as $\mu_p
= -\frac 3 2 \mu_n$ or the relative closeness of $N$ and $\Delta$
masses, are actually consequences of the $1/N_c$
expansion~\cite{DJM1,DJM2,Jenk}.  Equivalently, assuming that the
baryon masses and the $\pi N$ axial-current coupling $g_A$ scale as
$N_c^1$ (as naturally arises in quark and Skyrme models) leads to a
degenerate ``{\bf 56}'' multiplet through an analysis using
``consistency relations'' in $\pi N$ scattering, which are obtained by
the imposition of unitarity order-by-order in $N_c$~\cite{DJM1}.

The ground-state band ``{\bf 56}'' decomposes into multiplets with
spins $S_B \! = \! \frac 1 2, \frac 3 2, \ldots , \frac{N_c}{2}$, with
corresponding SU(3) representations [in the Dynkin weight notation
$(p,q)$] $(2S_B, \frac{N_c}{2} \! - \! S_B)$.  The first two members
of this series are denoted, as expected, ``{\bf 8}'' and ``{\bf 10}'',
while we may label the $S_B^{P_B} \! = \! {\frac 5 2}^+$, ${\frac 7
2}^+$, {\it etc.}\ members (those that disappear as $N_c \! \to \! 3$)
as ``large-$N_c$ exotic.''  The mass splittings between the multiplets
with $J \! = \!  O(N_c^0)$ are only $O(\Lambda_{\rm QCD}/N_c)$, which
for sufficiently large $N_c$ are smaller than $m_\pi$; hence, such
members of the ground-state multiplet are stable against strong
interactions and have widths that vanish in the large $N_c$ limit.
One may argue that it is a fluke of our universe that the chiral limit
is more closely realized than the large $N_c$ limit, allowing the
decay $\Delta \! \to \! \pi N$.

Once baryon states are determined to be stable in this way, they may
be analyzed using a Hamiltonian formalism, in which the spin-flavor
symmetry is broken perturbatively in powers of $1/N_c$ by operators
with specific quantum numbers under spin and flavor.  This method has
a long history and been dubbed the ``operator approach''~\cite{CL}.

Baryon resonances, on the other hand, are completely different
entities.  Appearing in meson-baryon scattering amplitudes, whose
generic size is $O(N_c^0)$, resonances arise as poles at complex
values of energy in the analytic continuation of these amplitudes.  Of
course, the real and imaginary part of each such value represents the
excitation mass and width, respectively, of the resonance, both of
which are typically $O(N_c^0)$.  Resonances are unstable against
strong decay even in the large $N_c$ limit, and require a treatment
distinct from that of the stable baryons.

Nevertheless, if a mechanism other than the $1/N_c$ expansion can be
invoked to suppress the creation of quark-antiquark pairs---the
mechanism through which a baryon resonance decays---then the treatment
of resonances as {\em almost\/} stable baryons becomes more
reasonable.  For example, if the quarks comprising the baryon are
heavy compared to $\Lambda_{\rm QCD}$ then pair creation is
suppressed, and an approach analogous to that used for the
ground-state baryons applies.  The analysis of ordinary baryon
resonances treated as almost stable in the $1/N_c$ expansion, using
the operator approach, has been carried out in great detail in the
literature~\cite{NStar_old}.

However, there exists a model-independent
treatment~\cite{CL,CLpenta} using the $1/N_c$ expansion in which the
resonances have natural [$O(N_c^0)$] widths and yet retains a good
deal of predictive power.  This ``scattering approach'' was
originally inspired~\cite{soliton} by the observation that numerous
results obtained in chiral soliton models such as the Skyrme model
appeared to be purely group-theoretical in origin.  A series of
papers in the 1980s by Mattis and collaborators~\cite{MP,MMSU2}
showed that such results are in fact independent of any dynamical
details of the models.  Indeed, in the case of two light quark
flavors the dominant $S$ matrix amplitudes were found to be
precisely those with t-channel exchange quantum numbers $I_t \! = \!
J_t$~\cite{MMSU2}.  This result can in turn be shown to follow
directly from the analysis of consistency relations derived from
scattering processes in large $N_c$~\cite{CL}.

Exploiting crossing relations, the $I_t \! = \! J_t$ rule can be
used to express observable meson-baryon scattering amplitudes in
terms of a smaller set of reduced amplitudes labeled in the s
channel by eigenvalues $K$ of the ``grand spin'' ${\bf K} \! = \!
{\bf I} \! + \! {\bf J}$.  In particular, a resonant pole appearing
in one scattering amplitude must appear in at least one of the
reduced amplitudes, which in turn appears in other scattering
amplitudes.  Baryon resonances therefore appear in multiplets
degenerate in both mass and width for large $N_c$.

The group theory for the three-flavor case relevant to hyperon
physics is of course more involved, but the necessary exercise was
carried out~\cite{MMSU3} by Mattis and Mukerjee in 1989 in the
context of soliton models. When three flavors are included, the
$I_t \! = \! J_t$ rule no longer holds in its original form, but as
we discuss below a more complicated set of constraints applies.  In
fact, in repeating the derivation of Ref.~\cite{MMSU3}, we find
small discrepancies, and discuss them below. Nevertheless, the
(suitably modified) Mattis-Mukerjee relation is the proper SU(3)
generalization of the SU(2) scattering relation at large $N_c$. It
predicts degenerate SU(3) multiplets of baryon resonances at large
$N_c$ and in the SU(3) limit. This observation and its practical
implementation (which required the computation of relevant SU(3)
Clebsch-Gordan coefficients (CGC)~\cite{CLSU3}) are the purposes of
this short paper.

As a first illustration of the power of the scattering method, we
consider the SU(3) partners to a hypothetical $J^P \! = \! \frac 1
2^+$ or $\frac 1 2^-$ isosinglet baryon resonance in an SU(3)
``$\overline{\bf 10}$''.  These are none other than two theoretically
favored sets of quantum numbers for the purported pentaquark state
$\Theta^+ (1540)$.  We discerned the partners of this $I \!  = \! 0$
state using the two-flavor formalism in Ref.~\cite{CLpenta}.  We of
course make no claims whether this state does indeed exist, but rather
conclude that if {\em any\/} baryon resonance with these quantum
numbers exists, then it must have partners degenerate in mass and
width at leading order in the $1/N_c$ expansion that carry specific
$J^P$ and SU(3) quantum numbers.

Moreover, we show below that the particular pattern of partners to
this state (for $P \! = \! +$) is precisely the one recently derived
by Jenkins and Manohar~\cite{JM04}.  They employed a $1/N_c$
operator approach originally inspired by the rigid rotor Skyrme
model, but consistent with the group theory of quark models as well.
One might have thought that such a coincidence is trivial.  However,
since the physical picture in Ref.~\cite{JM04} is based upon stable
states at large $N_c$, one can easily imagine the possibility that
when the widths of the states become order unity, mixing between the
multiplets given in Ref.~\cite{JM04} could occur.  This is not
unheard of in the scattering approach; indeed, the old
nonrelativistic SU(2$N_f$)$\times$O(3) quark model multiplets were
shown~\cite{CL} to form reducible collections of complete distinct
multiplets in the scattering approach.  It is therefore quite
heartening that the two methods agree so well in this case.

This paper is organized as follows.  In Sec.~\ref{amp}, we present the
master expression for three-flavor meson-baryon scattering and explain
its origin and relation to previous work.  Section~\ref{example}
presents a specific example: the enumeration of quantum numbers of
resonances degenerate in the large $N_c$ limit with a hypothetical
isoscalar, strangeness +1 resonance in the SU(3) ``$\overline{\bf
10}$'' representation, for the $J^P$ values ${\frac{1}{2}}^-$ and
${\frac{1}{2}}^+$.  These are of course the favored theoretical
preferences for the purported pentaquark $\Theta^+ (1540)$ state;
however, even if this state should turn out not to survive the current
experimental scrutiny, the example presented here should be viewed as
an indication of the power of the method.  In Sec.~\ref{concl} we
summarize, indicate future directions of research, and conclude.

\section{The SU(3) Amplitude Relation} \label{amp}

We now present the expression for the $S$-matrix amplitude in the
meson-baryon scattering process $\phi (S_\phi, R_\phi, I_\phi, Y_\phi)
+ B (S_B, R_B, I_B, Y_B) \to \phi^\prime (S_{\phi^\prime},
R_{\phi^\prime}, I_{\phi^\prime}, Y_{\phi^\prime}) + B^\prime
(S_{B^\prime}, R_{B^\prime}, I_{B^\prime}, Y_{B^\prime})$, where $S$,
$R$, $I$, and $Y$ stand, respectively, for the spin, SU(3)
representation, isospin, and hypercharge of the mesons $\phi$ and
$\phi^\prime$ and the baryons $B$ and $B^\prime$.  Primes here
indicate final-state quantum numbers.  The total {\em spin\/} angular
momentum (added vectorially) among the meson-baryon pairs are denoted
by $S$ and $S^\prime$, and the relative angular momenta between the
meson-baryon pairs are denoted by $L$ and $L^\prime$.  The amplitude
is described in terms of s-channel angular momentum $J_s$, SU(3)
representation $R_s$, isospin $I_s$, and hypercharge $Y_s$.  In
addition, multiple copies of $R_s$ can arise in the products $R_B
\otimes R_\phi$ and $R_{B^\prime} \otimes R_{\phi^\prime}$, and the
quantum numbers defined to lift this degeneracy are labeled by
$\gamma_s$ and $\gamma^\prime_s$ (which need not be equal).  That the
other s-channel quantities are conserved can be demonstrated
explicitly ({\it e.g.}, $R_s \! = \!  R^\prime_s$), and thus the
primes on such quantities are suppressed.  The amplitude is reduced,
in the sense of the Wigner-Eckart theorem, in that the SU(2) quantum
numbers $J_{sz}$ and $I_z$ do not appear explicitly.  The notation
$[X]$ refers to the dimension of a given representation, whether $X$
is labeled by $I$ or $J$ in SU(2), or by the actual dimension in SU(3)
({\it i.e.}, [$J \! = \! 1$] = 3, but [$R \! = \! {\bf 8}$] = 8).  The
master expression for such scattering amplitudes in the large-$N_c$
limit then reads
\begin{eqnarray}
\lefteqn{S_{L L^\prime S S^\prime J_s R_s \gamma_s \gamma^\prime_s I_s
Y_s}} \nonumber \\
& = & (-1)^{S_B - S_{B^\prime}}
([R_B][R_B^\prime][S][S^\prime])^{1/2} / [R_s]
\sum_{\stackrel{\scriptstyle I \in R_\phi, \; I^\prime \in
R_{\phi^\prime},}{I^{\prime\prime} \in R_s, \; Y \in R_\phi \cap
R_{\phi^\prime}}} (-1)^{I + I^\prime + Y} [I^{\prime\prime}]
\nonumber \\ & & \times
\left( \begin{array}{cc||c} R_B & R_\phi & R_s \, \gamma_s \\ S_B
\frac{N_c}{3} & I Y & I^{\prime\prime} \, Y \! \! + \! \frac{N_c}{3}
\end{array} \right)
\left( \begin{array}{cc||c} R_B & R_\phi & R_s \, \gamma_s \\ I_B
Y_B & I_\phi Y_\phi & I_s Y_s \end{array} \right) \nonumber \\
& & \times
\left( \begin{array}{cc||c} R_{B^\prime} & R_{\phi^\prime} & R_s \,
\gamma^\prime_s \\ S_{B^\prime}
\frac{N_c}{3} & I^\prime Y & I^{\prime\prime} \, Y \! \! + \!
\frac{N_c}{3}
\end{array} \right)
\left( \begin{array}{cc||c} R_{B^\prime} & R_{\phi^\prime} & R_s \,
\gamma^\prime_s \\ I_{B^\prime} Y_{B^\prime} & I_{\phi^\prime}
Y_{\phi^\prime} & I_s Y_s \end{array} \right)
\nonumber \\ & & \times \sum_{K, \tilde{K} , \tilde{K}^\prime}
[K] ([\tilde{K}][\tilde{K}^\prime])^{1/2}
\left\{ \begin{array}{ccc}
L   & I                & \tilde{K} \\
S   & S_B              & S_\phi    \\
J_s & I^{\prime\prime} & K \end{array} \right\}
\! \left\{ \begin{array}{ccc}
L^\prime & I^\prime         & \tilde{K}^\prime \\
S^\prime & S_{B^\prime}     & S_{\phi^\prime} \\
J_s      & I^{\prime\prime} & K \end{array} \right\}
\tau^{\left\{ I I^\prime Y \right\}}_{K \tilde{K} \tilde{K}^\prime \!
L L^\prime} \ .
\label{Mmaster}
\end{eqnarray}
The quantities containing double vertical bars are SU(3) isoscalar
CGC~\cite{CLSU3}, while those in braces are ordinary SU(2) $9j$
symbols.  This expression should be compared with the original
Mattis-Mukerjee result [Ref.~\cite{MMSU3} Eq.~(12)].  Since its
derivation was a primary result of that paper, we do not present a
detailed rederivation here, but merely discuss its structure, and
then detail differences between the present expression and that of
Ref.~\cite{MMSU3} (in particular, why expression in
Ref.~\cite{MMSU3} is not suitable for physical processes).

The two-flavor scattering formula is derived in the earlier chiral
soliton-type treatment~\cite{MP} by starting with a fundamental
soliton in the conventional hedgehog configuration, which is an
eigenstate of the grand spin ${\bf K} \! \equiv \! {\bf I} \! + {\bf
J}$.  Scattering is accomplished by the standard linear expansion of
the soliton in terms of pion field fluctuations.  However, physical
hadrons are of course specified not by $K$ but by $I$ and $J$, and
hence one must allow for multiple values of $K$ in a full physical
scattering process order to form a linear superposition that is an
eigenstate of $I$ and $J$; nevertheless, one may treat $K$ as a
hidden degree of freedom conserved in the underlying scattering
processes, which therefore attaches as a label to the reduced
scattering amplitudes $\tau$.  In generalizing the process to allow
for mesons $\phi$, $\phi^\prime$ of arbitrary isospin $I$ and spin
$S$, one requires also the intermediate quantum numbers $\tilde {\bf
K} \! \equiv \! {\bf I}_\phi \! + \! {\bf L}$ and $\tilde {\bf
K}^\prime \! \equiv \!  {\bf I}_{\phi^\prime} \!  + \!  {\bf
L}^\prime$ (so that ${\bf K} \! = \! \tilde {\bf K} \! + \!  {\bf
S}_\phi \! = \! \tilde {\bf K}^\prime \!  + \! {\bf
S}_{\phi^\prime}$) used in Eq.~(\ref{Mmaster}).  The $9j$ symbols
simply arise through the combination of the numerous SU(2) CGC that
arise in this procedure from the vector addition of multiple
SU(2)-valued quantities.

The three-flavor generalization is conceptually quite
straightforward, if mathematically more cumbersome: One simply
rotates the full initial and final states into their nonstrange
partners in the same irreducible SU(3) representation, and uses the
two-flavor expression for the nonstrange scattering process.  The
inclusion of SU(3) rotation matrices of course introduces the SU(3)
CGC in Eq.~(\ref{Mmaster}).

In repeating the derivation of Ref.~\cite{MMSU3} Eq.~(12) to obtain
our Eq.~(\ref{Mmaster}), we find a few small but significant
differences. First, the overall phase of the original result lacks
our phase $(-1)^{S_B \! - \! S_{B^\prime}}$.  Second,
Ref.~\cite{MMSU3} appears to average over baryons and mesons in the
external states with all possible quantum numbers within the given
SU(3) multiplets (rendering their expression phenomenologically less
useful); if we do the same with Eq.~(\ref{Mmaster}), two of our
SU(3) CGC are absorbed through an orthogonality relation, matching
the older result.  Finally, their explicit unity values for the
nonstrange baryon hypercharges must be modified to $\frac{N_c}{3}$,
in light of the proper quantization~\cite{WZ} of the Wess-Zumino
term for arbitrary $N_c$.  While these agree for $N_c \! = \! 3$, it is
important to keep the general form so that consistency in $N_c$ scaling
can be verified.

A difference in our interpretation relative to Ref.~\cite{MMSU3} also
helps resolve a paradox of that work.  In Ref.~\cite{MMSU3} it was
noted that the $I_t \! = \! J_t$ rule does not hold for meson-baryon
scattering in soliton models with SU(3) flavor at leading order in
$N_c$, even for processes with no exchange of strangeness.  This is
worrying, since as noted in Refs.~\cite{CL,CLpenta} the $I_t \! = \!
J_t$ rule can be derived for such processes directly from large $N_c$
QCD with no additional model assumptions.  The origin of this
perplexing discrepancy is that the SU(3) representations used for the
baryons of interest in Ref.~\cite{MMSU3} are those which occur for
$N_c \! = \!  3$ ({\it e.g.}, the literal {\bf 8} and {\bf 10}).
However, as noted in the Introduction, the appropriate representations
in a large $N_c$ world are not these but rather the ``{\bf 8}'',
``{\bf 10}'', and so on.  Strictly speaking the relations derived here
hold for the large $N_c$ world, and thus one expects the $I_t \! = \!
J_t$ rule to hold for meson-baryon scattering with strangeness and the
baryons in their large $N_c$ representations.  Thus, by using the $N_c
\! = \! 3$ representations, Ref.~\cite{MMSU3} implicitly includes
specific $1/N_c$ corrections.  However, since this was done with $N_c$
set to 3, one could not cleanly isolate the numerically small
violation of the $I_t \! = \! J_t$ rule as a $1/N_c$ correction.
Using the proper large $N_c$ representations and formulae derived
here, the $I_t \! = \! J_t$ rule indeed holds.

For phenomenological purposes, the most interesting special case of
Eq.~(\ref{Mmaster}) is that in which the baryons belong to the
parity-positive ground-state ``{\bf 56}'' multiplet, and the mesons
are both pseudoscalar SU(3)-octet pseudo-Nambu-Goldstone bosons.  In
this case, parity ($P$) conservation demands that $L \! - \! L^\prime$
is an even integer, and the $9j$ symbols collapse to $6j$ symbols.
The master scattering amplitude for the process $\phi (S_\phi \! = \!
0, R_\phi \!  = \! 8, I_\phi, Y_\phi) + B (S_B, R_B, I_B, Y_B) \to
\phi^\prime (S_{\phi^\prime} \! = \! 0, R_{\phi^\prime} \! = \! 8,
I_{\phi^\prime}, Y_{\phi^\prime}) + B^\prime (S_{B^\prime},
R_{B^\prime}, I_{B^\prime}, Y_{B^\prime})$ then reads
\begin{eqnarray}
\lefteqn{S_{L L^\prime S_B S_{B^\prime} J_s R_s \gamma_s
\gamma^\prime_s I_s Y_s}} \nonumber \\
& = & (-1)^{S_B - S_{B^\prime}}
([R_B][R_B^\prime])^{1/2} / [R_s]
\sum_{\stackrel{\scriptstyle I, I^\prime \! , \, Y \in
8,}{I^{\prime\prime} \in R_s}}
(-1)^{I + I^\prime + Y} [I^{\prime\prime}]
\nonumber \\ & & \times
\left( \begin{array}{cc||c} R_B & 8 & R_s \, \gamma_s \\ S_B
\frac{N_c}{3} & I Y & I^{\prime\prime} \, Y \! \! + \! \frac{N_c}{3}
\end{array} \right)
\left( \begin{array}{cc||c} R_B & 8 & R_s \, \gamma_s \\ I_B
Y_B & I_\phi Y_\phi & I_s Y_s \end{array} \right) \nonumber \\
& & \times
\left( \begin{array}{cc||c} R_{B^\prime} & 8 & R_s \,
\gamma^\prime_s \\ S_{B^\prime}
\frac{N_c}{3} & I^\prime Y & I^{\prime\prime} \, Y \! \! + \!
\frac{N_c}{3} \end{array} \right)
\left( \begin{array}{cc||c} R_{B^\prime} & 8 & R_s \,
\gamma^\prime_s \\ I_{B^\prime} Y_{B^\prime} & I_{\phi^\prime}
Y_{\phi^\prime} & I_s Y_s \end{array} \right)
\nonumber \\ & & \times \sum_{K} [K]
\left\{ \begin{array}{ccc}
K   & I^{\prime\prime} & J_s \\
S_B & L                & I \end{array} \right\}
\! \left\{ \begin{array}{ccc}
K            & I^{\prime\prime} & J_s \\
S_{B^\prime} & L^\prime         & I^\prime \end{array} \right\}
\tau^{\left\{ I I^\prime Y \right\}}_{K K K L L^\prime} \ .
\label{spinless}
\end{eqnarray}

While these expressions were originally derived within the context
of a chiral soliton picture, they are model-independent consequences
of QCD in the large $N_c$ limit.  By employing crossing relations,
one first shows~\cite{MMSU2} that the conservation of $K$ in
$s$-channel processes for the two-flavor case is equivalent to the
rule $I_t \! = \! J_t$ in the $t$ channel; and as discussed in
Sec.~\ref{intro}, this rule is a direct large $N_c$
consequence~\cite{CL}.  The three-flavor generalization, in turn, is
simply an SU(3) rotation of the SU(2) result with no additional
dynamics, and therefore is also a model-independent large $N_c$
result.

\section{Explicit Example} \label{example}

To illustrate the utility of the approach, we apply it to find SU(3)
partners of the reported narrow $\Theta^+$ exotic pentaquark
resonance.  A few caveats are useful before proceeding.  First, there
is considerable controversy as to whether these states are real; in
this work we take an agnostic position.  The issue addressed here is
that {\em if\/} the resonance is real, then at large $N_c$ and in the
SU(3) limit it must have degenerate partners, which at finite $N_c$
would correspond to some {\em nearly\/} degenerate partners; our goal
is to enumerate them.  Second, the analysis is based on exact SU(3)
symmetry.  While small SU(3) violations can be accounted for
perturbatively, if for some reason large SU(3) violations occur then
the present formalism breaks down.  For example, it has been suggested
by Jaffe and Wilczek~\cite{JW} in the context of a diquark model that
nearly ideal mixing may occur between different SU(3) multiplets (such
as for $\phi$ and $\rho$ mesons), which could lead to large SU(3)
violations.  The present work assumes that such a scenario does not
occur; indeed, sound phenomenological arguments~\cite{Coh} oppose it.
Third, if the $\Theta^+$ does in fact exist, then apart from its
strangeness and isospin we do not know its quantum numbers directly
from experiment.  Since predictions of SU(3) partners depend upon the
quantum numbers, we assume here that the $\Theta^+$ is a spin-$\frac 1
2$ isoscalar, which seems to be the most natural possibility from a
theoretical perspective.  The parity of the $\Theta^+$ is unknown, and
various plausible theoretical arguments can be made to suggest either
parity; accordingly, we consider both cases.  Finally, it remains
possible that the $\Theta^+$ exists and has different quantum numbers.
In such a case one could perform an analysis entirely analogous to the
one considered here.

\subsection{``Seed'' Amplitudes}

We suppose that a pole corresponding to a baryon resonance appears
in an $NK$ partial wave ($N$: $S_B \! = \! S_{B^\prime} \! = \!
\frac 1 2$, $P_B \! = \! P_{B^\prime} \! = +$, $I_B \! = \!
I_{B^\prime} \! = \! \frac 1 2$, $Y_B \! = \!  Y_{B^\prime} \! = \!
\frac{N_c}{3}$, $R_B \! = \! R_{B^\prime} \! = \!  {\rm ``} {\bf
8}$''; $K$: $I_\phi \! = \! I_{\phi^\prime} \! = \! \frac 1 2$,
$Y_\phi \! = \! Y_{\phi^\prime} \! = \! 1$, $R_{\phi} \! = \!
R_{\phi^\prime} \! = \! {\bf 8})$ with quantum numbers $I_s \! = \!
0$, $Y_s \! = \! \frac{N_c}{3} \! + \! 1$, $J_s \! = \! \frac 1 2$,
$R_s \! = \!  {\rm ``} \overline{\bf 10}$'', and examine the
consequences of Eq.~(\ref{spinless}).   The first task is to
determine which reduced amplitudes $\tau$ contribute to partial
waves carrying these quantum numbers, and therefore act as ``seed''
amplitudes to produce poles in other partial waves.  As we now show,
only $\tau^{\{ \frac 1 2 \frac 1 2 1 \} }_{\frac 1 2 \frac 1 2 \frac
1 2 LL}$ appears (with $L \! = \!  0,1$ for $P_s \! = \! \mp$),
implying that the assumed resonant pole must lie in that reduced
amplitude; had several amplitudes arisen, it would have necessary to
perform a more delicate analysis to look for degenerate poles in
multiple partial waves in order to determine which reduced
amplitudes they have in common.

The triangle rules imposed by Eq.~(\ref{spinless}) are $\delta ( S_B I
I^{\prime\prime} )$, $\delta ( S_{B^\prime} I^\prime I^{\prime\prime}
)$, $\delta ( K I^{\prime\prime} J_s )$, $\delta ( K L I )$, $\delta (
K L^\prime I^\prime )$, $\delta( S_B L J_s )$, and $\delta (
S_{B^\prime} L^\prime J_s )$.  Imposing the substitutions listed
above, the last two imply $L \! = \! L^\prime \! = \! 0$ or 1 for $P_s
\! = \! -$ or $+$, respectively.  If $L \! = \! L^\prime \! = \! 0$,
then $I^{\prime\prime}$ equals either $\pm \frac 1 2$ added to the
common value $I \! = \! I^\prime \! = \! K$.  On the other hand, if $L
\! = \! L^\prime \! = \! 1$, then satisfying the triangle rules
requires that each of $I$, $I^\prime$, and $K$ differ from
$I^{\prime\prime}$ by $+\frac 1 2$ or $-\frac 1 2$.  The sums in
Eq.~(\ref{spinless}) are truncated by the requirement that $(I,Y)$ and
($I^\prime,Y)$ must be the quantum numbers of states within a literal
SU(3) octet, which are $(\frac 1 2 , 1)$, $(1,0)$, $(0,0)$, and
$(\frac 1 2 , -1)$.

The current case is simplified considerably by noting that
``$\overline{\bf 10}$'' = [0, $(N_c \! + 3)/2$] contains only
singly-degenerate states; in particular, one finds using the variables
of Eq.~(\ref{spinless}) that $2I^{\prime\prime} \!  + Y
\! = \! 1$.  To each isospin multiplet $(I,Y)$ or $(I^\prime , Y)$
within the {\bf 8} one therefore identifies a unique value
$I^{\prime\prime} \! = \! (1 \! - \! Y)/2$.  The required SU(3) CGC
then assume the form
\begin{equation}
\left( \begin{array}{cc||c} ``8$''$ & 8 & ``\overline{10}$''$
\\ \frac 1 2 , \frac{N_c}{3} & \{ I, I^\prime \} , Y & \frac{1 - Y}{2}
, \, Y \! \! + \! \frac{N_c}{3}
\end{array} \right) \ .
\end{equation}
All of these CGC are compiled in Table~I of Ref.~\cite{CLSU3}.  From
this source one readily determines that the only such coefficient
nonvanishing in the large $N_c$ limit has $( \{ I, I^\prime \} ,Y) \! =
\! ( \frac 1 2 , 1 )$, for which the CGC equals $-1$.  But then also
$I^{\prime\prime} \! = \! 0$, which forces not only $I \! = \!
I^\prime \! = \! \frac 1 2$ and $Y \! = \! 1$, but also $K \! = \!
\frac 1 2$.  It follows that the unique reduced amplitude contributing
in the each of the $L \! = \! 0$ and $L \! = \! 1$ cases is, as
promised, $\tau^{\{ \frac 1 2 \frac 1 2 1 \} }_{\frac 1 2 \frac 1 2
\frac 1 2 LL}$.

To complete the simplification of Eq.~(\ref{spinless}) for this case,
we note that $[K] \! = \! 2$, the dimension of any SU(3)
representation $(p,q)$ assumes the usual value $\frac 1 2 (p \! + \!
1)(q \! + \! 1)(p \! + \! q \!  + \! 2)$: [``{\bf 8}''] = $\frac 1 4
(N_c \! + \! 5)(N_c \! + \! 1)$ and [``$\overline{\bf 10}$''] = $\frac
1 8 (N_c \! + \! 7)(N_c \! + \! 5)$, and the $6j$ symbols give
\begin{equation}
\left\{ \begin{array}{ccc}
\frac 1 2 & 0         & \frac 1 2 \\
\frac 1 2 & \{ 0,1 \} & \frac 1 2
\end{array} \right\} = \mp \frac 1 2  \ .
\end{equation}
In the large $N_c$ limit, one then finds the numerical coefficient
of $\tau^{\{ \frac 1 2 \frac 1 2 1 \} }_{\frac 1 2 \frac 1 2 \frac 1
2 LL}$ to be unity.

\subsection{Degrees of ``Exoticness''}

Before continuing, it is important to point out that the concept of
``exoticness'' can be used in three different but related senses
here: First, we label the members of the ``{\bf 56}'' with $S_B \! >
\! \frac 3 2$ as ``large-$N_c$ exotic'' because their corresponding
SU(3) representations $(2S_B, \frac{N_c}{2} - S_B)$ are not allowed
for $N_c \! = \!  3$.  We denote processes in which the initial
ground-state baryon is large-$N_c$ exotic by ${\cal E}$, nonexotic
by ${\cal N}$.  Second, the product representation in which the
baryon resonances appear may be nonexotic (${\cal N}^*$) or exotic
in one of two distinct ways: Either it is a perfectly ordinary SU(3)
representation that cannot be produced through a $qqq$ state, such
as the ``$\overline{\bf 10}$'' (denoted by ${\cal E}^*_0$), or it
may also be large-$N_c$ exotic (denoted by ${\cal E}^*_1$).  All 6
types of scattering process, ${\cal N} {\cal N}^*$, ${\cal N} {\cal
E}^*_0$, ${\cal N} {\cal E}^*_1$, ${\cal E} {\cal N}^*$, ${\cal E}
{\cal E}^*_0$, and ${\cal E} {\cal E}^*_1$, may occur in a large
$N_c$ world.

The possibility of some of the mixed ${\cal N}$-${\cal E}$
combinations may come as a bit of a surprise.  As an example of an
${\cal N} {\cal E}^*_1$ process, note that the product ``{\bf
8}''$\otimes \, {\bf 8}$ contains the large-$N_c$ exotic SU(3)
representation $[2, (N_c - \! 5)/2]$.  On the other hand, ${\cal E}
{\cal N}^*$ also can occur: The ${\frac 5 2}^+$ ground-state baryon
can scatter a pseudoscalar {\bf 8} meson to give a ``{\bf 10}''
resonance.  For our present purposes, we are interested in ${\cal
E}^*_0$ processes, both ``singly exotic'' (${\cal N} {\cal E}^*_0$)
and ``doubly exotic'' (${\cal E} {\cal E}^*_0$).  That is, we are
interested in exotic resonances that lie in SU(3) representations
existing at $N_c \! = 3$, but allow for the possibility that the
ground-state baryon representation needed to produce them in
scattering with a pseudoscalar {\bf 8} meson might itself not occur
for $N_c \! = 3$.  We make this choice to mirror the terminology of
Ref.~\cite{JM04}.

One can readily show that there exists an upper limit to ground-state
baryon spins $S_B$ in ``{\bf 56}'' allowing ${\cal E} {\cal E}^*_0$
processes via scattering with {\bf 8} mesons (beyond which only ${\cal
E} {\cal E}^*_1$ occurs).  As $S_B$ increases, the second row of its
SU(3) tableau (length $\frac{N_c}{2} \! - \! S_B$) becomes so short
that the 3 boxes in the {\bf 8} are insufficient to produce a
large-$N_c$ nonexotic resonance.  By direct computation, one finds
that the ${\cal N} {\cal E}^*_0$ possibilities are ``{\bf 8}''$\otimes
\,${\bf 8}$\to$``$\overline{\bf 10}$'' and ``{\bf 27}'', ``{\bf
10}''$\otimes \,${\bf 8}$\to$``{\bf 27}'' and ''{\bf 35}'' (where
``{\bf 27}'' = $[2, (N_c + 1)/2]$ and ``{\bf 35}'' = $[4, (N_c - \!
1)/2]$), and the ${\cal E} {\cal E}^*_0$ possibilities are ${\frac 5
2}^+ \! \otimes {\bf 8} \! \to$``{\bf 28}'' and ``{\bf 35}'' (where
``{\bf 28}'' = $[6, (N_c -  3)/2]$), and ${\frac 7 2}^+
\! \otimes {\bf 8} \! \to$``{\bf 28}''.

\subsection{Finding SU(3) Partners}

Having isolated the reduced amplitudes containing the desired resonant
pole, we now reverse the process in order to determine the full set of
partial waves to which these reduced amplitudes contribute.  The
triangle rules imposed by Eq.~(\ref{spinless}) with $I \! = \!
I^\prime \! = \! K \! = \! \frac 1 2$ force each of $L$, $L^\prime$ to
equal either 0 or 1; and the fact that all baryons in the ground-state
``{\bf 56}'' have $P_B \! = \! +$ again forces $L \! = \! L^\prime$.
Note in particular that this procedure obtains only degenerate
partners all carrying the same parity.

\subsubsection{Negative Parity}

We first analyze the case $P_s \! = \! -$ case, for which $L \! = \!
L^\prime \! = \! 0$.  Then $S_B \! = \! S_{B^\prime} \! = \! J_s$, so
that $R_B \! = \! R_{B^\prime}$, and the only remaining nontrivial
triangle rule is $\delta ( S_B \frac 1 2 I^{\prime\prime} )$;
Eq.~(\ref{spinless}) collapses to
\begin{eqnarray}
S_{0 0 S_B S_{B^\prime} J_s R_s \gamma_s
\gamma^\prime_s I_s Y_s}
& = &
\delta_{R_B R_{B^\prime}} \delta_{S_B S_{B^\prime}} \delta_{S_B J_s}
\frac{[R_B]}{[R_s] [S_B]}
\, \tau^{\left\{ \frac{1}{2} \frac{1}{2} 1 \right\}}_{\frac{1}{2}
\frac{1}{2} \frac{1}{2} 0 0}
\sum_{I^{\prime\prime} \in R_s} [I^{\prime\prime}]
\nonumber \\ & & \times
\left( \begin{array}{cc||c} R_B & 8 & R_s \, \gamma_s \\ S_B \,
\frac{N_c}{3} & \frac{1}{2} \, 1 & I^{\prime\prime} \,
\frac{N_c}{3} \! + \! 1
\end{array} \right)
\left( \begin{array}{cc||c} R_B & 8 & R_s \, \gamma_s \\ I_B
Y_B & I_\phi Y_\phi & I_s Y_s \end{array} \right) \nonumber \\ & &
\times
\left( \begin{array}{cc||c} R_B & 8 & R_s \,
\gamma^\prime_s \\ S_B \,
\frac{N_c}{3} & \frac{1}{2} \, 1 & I^{\prime\prime} \,
\frac{N_c}{3} \! + \! 1 \end{array} \right)
\left( \begin{array}{cc||c} R_{B^\prime} & 8 & R_s \,
\gamma^\prime_s \\ I_{B^\prime} Y_{B^\prime} & I_{\phi^\prime}
Y_{\phi^\prime} & I_s Y_s \end{array} \right) \ .
\end{eqnarray}
In order to study only ${\cal E}^*_0$ processes, as discussed above we
limit $S_B \! \le \frac 7 2$.  The CGC for $S_B \! = \! \frac 1 2$
($R_B \! = $``{\bf 8}'') and $\frac 3 2$ ($R_B \! = $``{\bf 10}'')
again all appear in Ref.~\cite{CLSU3}.  One finds that the only CGC
surviving as $N_c \! \to \! \infty$ for $S_B \! = \! \frac 1 2$ have
either $R_s$ = ``$\overline{\bf 10}$'', $I^{\prime\prime} \! = \! 0$
(giving $-1$), or $R_s$ = ``{\bf 27}'', $I^{\prime\prime} \! = \! 1$
($+1$).  For $S_B \! = \! \frac 3 2$, we have either $R_s$ = ``{\bf
27}'', $I^{\prime\prime} \!  = \! 1$ ($-1$), or $R_s$ = ``{\bf 35}'',
$I^{\prime\prime} \! = \! 2$ ($+1$).  Reference~\cite{CLSU3} does not
compile CGC for $S_B \! = \! \frac 5 2$ or $\frac 7 2$ baryons, but
for our purposes it is only necessary to know that there exist
$O(N_c^0)$ couplings for $S_B \! = \! \frac 5 2$ to $R_s$ = ``{\bf
35}'' and ``{\bf 28}'', and for $S_B \! = \! \frac 7 2$ to $R_s$ =
``{\bf 28}'' and the ${\cal E}^*_1$ representation $[8, (N_c \! - \!
5)/2]$.  Indeed, for the states of maximal hypercharge in $R_B$
($\frac{N_c}{3}$), {\bf 8} ($+1$), and $R_s$ ($\frac{N_c}{3} \! + \!
1$) [as required by the first CGC in Eq.~(\ref{spinless})], it is
straightforward to show that only one SU(3) representation occurs for
$I^{\prime\prime} \! = \! S_B + \! \frac 1 2$ [$(p,q) \! = \!  (2S_B +
\! 1, \frac{N_c}{2} \! - S_B \! + \! 1)$], and only one occurs for
$I^{\prime\prime} \! = \! S_B \! - \! \frac 1 2$ [$(p,q) \! = \!
(2S_B \! - \! 1, \frac{N_c}{2} \! - S_B \! + \! 2)$], which are the
representations listed above.  The CGC in each of these cases must
therefore be either $+1$ or $-1$.

Collecting these results, one then finds the set of degenerate
multiplets $(R_s, J_s^-)$ to be (``$\overline{\bf 10}$'', ${\frac 1
2}^-$), (``{\bf 27}'', ${\frac 1 2}^-$), (``{\bf 27}'', ${\frac 3
2}^-$), and (``{\bf 35}'', ${\frac 3 2}^-$) (singly exotic, via ${\cal
N} {\cal E}^*_0$ processes), and (``{\bf 35}'', ${\frac 5 2}^-$),
(``{\bf 28}'', ${\frac 5 2}^-$), and (``{\bf 28}'', ${\frac 7 2}^-$)
(doubly exotic, via ${\cal E} {\cal E}^*_0$ processes).

\subsubsection{Positive Parity}

The case $P_s \! = \! +$, for which $L \! = \! L^\prime \! = \! 1$, is
only a bit more complicated.  Now one may have $S_B \! \neq \!
S_{B^\prime}$ and $R_B \! \neq \! R_{B^\prime}$, and $J_s$ must be
separately specified.  In this case, Eq.~(\ref{spinless}) becomes
\begin{eqnarray}
S_{1 1 S_B S_{B^\prime} J_s R_s \gamma_s \gamma^\prime_s I_s Y_s}
& = & (-1)^{S_B - S_{B^\prime}}
\, \frac{2 ([R_B][R_B^\prime])^{1/2}}{[R_s]} \,
\tau^{\left\{ \frac{1}{2} \frac{1}{2} 1 \right\}}_{\frac{1}{2}
\frac{1}{2} \frac{1}{2} 1 1}
\sum_{I^{\prime\prime} \in R_s}
[I^{\prime\prime}]
\nonumber \\ & & \times
\left( \begin{array}{cc||c} R_B & 8 & R_s \, \gamma_s \\ S_B
\frac{N_c}{3} & \frac 1 2 \, 1 & I^{\prime\prime} \,
\frac{N_c}{3} \! + \! 1
\end{array} \right)
\left( \begin{array}{cc||c} R_B & 8 & R_s \, \gamma_s \\ I_B
Y_B & I_\phi Y_\phi & I_s Y_s \end{array} \right) \nonumber \\
& & \times
\left( \begin{array}{cc||c} R_{B^\prime} & 8 & R_s \,
\gamma^\prime_s \\ S_{B^\prime}
\frac{N_c}{3} & \frac 1 2 \, 1 & I^{\prime\prime} \,
\frac{N_c}{3} \! + \! 1
\end{array} \right)
\left( \begin{array}{cc||c} R_{B^\prime} & 8 & R_s \,
\gamma^\prime_s \\ I_{B^\prime} Y_{B^\prime} & I_{\phi^\prime}
Y_{\phi^\prime} & I_s Y_s \end{array} \right)
\nonumber \\ & & \times
\left\{ \begin{array}{ccc}
\frac 1 2 & I^{\prime\prime} & J_s \\
S_B       & 1                & \frac 1 2 \end{array} \right\}
\! \left\{ \begin{array}{ccc}
\frac 1 2    & I^{\prime\prime} & J_s \\
S_{B^\prime} & 1                & \frac 1 2 \end{array} \right\} \ .
\end{eqnarray}
Note that precisely the same set of SU(3) CGC are relevant to the case
$P_s \! = \! +$, meaning that the enumeration of SU(3) representations
carries over verbatim from the case $P_s \! = \! -$; only the angular
momenta need be considered more carefully.  The remaining independent
triangle rules imposed by the $6j$ symbols are $\delta ( S_B \, \frac
1 2 \, I^{\prime\prime} )$, $\delta ( S_{B^\prime} \, \frac 1 2 \,
I^{\prime\prime} )$, and $\delta ( \frac 1 2 \, I^{\prime\prime} J_s
)$.  One then finds the following combinations.  From $S_B \! = \!
\frac 1 2, R_B \! =$``{\bf 8}'': $I^{\prime\prime} \! = \! 0 \to R_s
\! =$``$\overline{\bf 10}$'', $J_s \! = \! \frac 1 2$ and
$I^{\prime\prime} \! = \! 1 \to R_s \! =$``{\bf 27}'', $J_s \! = \!
\frac{1}{2}, \! \frac{3}{2}$.  From $S_B \! = \! \frac 3 2, R_B \!
=$``{\bf 10}'': $I^{\prime\prime} \! = \! 1 \to R_s \! =$``{\bf 27}'',
$J_s \! = \! \frac{1}{2}, \! \frac{3}{2}$ and $I^{\prime\prime} \! =
\! 2 \to R_s \! =$``{\bf 35}'', $J_s \! = \! \frac{3}{2}, \!
\frac{5}{2}$.  From $S_B \! = \! \frac 5 2, R_B \! = \! [5, (N_c \! -
\! 5)/2]$: $I^{\prime\prime} \! = \! 2 \to R_s \! =$``{\bf 35}'', $J_s
\! = \! \frac{3}{2}, \! \frac{5}{2}$ and $I^{\prime\prime} \! = \! 3
\to R_s \! =$``{\bf 28}'', $J_s \! = \! \frac{5}{2}, \! \frac{7}{2}$.
And from $S_B \! = \! \frac 7 2, R_B \! = \! [7, (N_c \! - \! 7)/2]$:
$I^{\prime\prime} \! = \! 3 \to R_s \! =$``{\bf 28}'', $J_s \! = \!
\frac{5}{2}, \! \frac{7}{2}$ and $I^{\prime\prime} \! = \! 4 \to R_s
\! = \! [8, (N_c \! - \! 5)/2]$, $J_s \! = \! \frac{7}{2}, \!
\frac{9}{2}$.

Collecting these results, one then finds the set of degenerate
multiplets $(R_s, J_s^+)$ to be (``$\overline{\bf 10}$'', ${\frac 1
2}^+$), (``{\bf 27}'', ${\frac 1 2}^+$), (``{\bf 27}'', ${\frac 3
2}^+$), (``{\bf 35}'', ${\frac 3 2}^+$), and (``{\bf 35}'', ${\frac 5
2}^+$) (singly exotic, via ${\cal N} {\cal E}^*_0$ processes), and
(``{\bf 28}'', ${\frac 5 2}^+$), and (``{\bf 28}'', ${\frac 7 2}^+$)
(doubly exotic, via ${\cal E} {\cal E}^*_0$ processes).  As promised,
these multiplets precisely match those obtained in Ref.~\cite{JM04}
via counting using Young tableaux, once a consistent definition of
degree of exoticness is included.

\section{Conclusions} \label{concl}

We have generalized to three flavors the two-flavor large $N_c$
meson-baryon scattering method that relates different scattering
partial waves.  In particular, resonant poles occurring in one such
amplitude appear in others, creating multiplets of resonances
degenerate in both mass and width at leading order in the $1/N_c$
expansion limit.  We illustrated the method by finding the partners of
a resonance carrying the quantum numbers suggested by the purported
$\Theta^+ (1540)$ particle, and found that the results (for $J^P
\! = \! \frac 1 2^+$) agree with those obtained using a large
$N_c$ method that does not recognize the instability of the
resonances.

One may immediately apply this formalism to numerous other problems
involving three-flavor baryon resonances.  For example, in it has been
shown~\cite{CL} that the suppression of the $N(1535)$ $\pi N$ partial
width is due to the fact that the reduced amplitude appearing in the
$S_{11}$ partial wave couples in the large $N_c$ limit does not couple
to spinless isovector mesons.  What interesting analogous consequences
arise for the $\Lambda$ and $\Xi$ resonances?

Finally, the $I_t \! = \! J_t$ rule was used~\cite{CL} to parametrize
$1/N_c$ corrections to the leading-order large $N_c$ results, an
absolute must for useful phenomenological studies.  As discussed
above, the crossing constraint for three flavors cannot be described
so simply.  We defer to future work the description and application of
this important concept.

{\it Acknowledgments.}  T.D.C.\ was supported by the D.O.E.\ through
grant DE-FGO2-93ER-40762; R.F.L.\ was supported by the N.S.F.\ through
grant PHY-0140362.

\end{document}